\documentclass[conference]{IEEEtran}
\IEEEoverridecommandlockouts
\usepackage{cite}
\usepackage{amsmath,amssymb,amsfonts}
\usepackage{algorithmic}
\usepackage{graphicx}
\usepackage{multirow}
\usepackage{verbatim}
\usepackage{textcomp}
\usepackage{xcolor}
\usepackage{makecell}
\usepackage{bm}
\usepackage{graphicx}

\usepackage{verbatim}
\usepackage{float}
\usepackage[caption=false,font=footnotesize,labelfont=rm,textfont=rm]{subfig}
\usepackage[ruled,linesnumbered]{algorithm2e}

\def\BibTeX{{\rm B\kern-.05em{\sc i\kern-.025em b}\kern-.08em
    T\kern-.1667em\lower.7ex\hbox{E}\kern-.125emX}}
\begin{document}

\title{
Deep Learning Enabled Semantic Communication Systems for Video Transmission\\
\thanks{This work is partly supported by NSFC under grant No. 62201505, partly by the SUTD-ZJU IDEA Grant (SUTD-ZJU (VP) 202102), and partly by the Fundamental Research Funds for the Central Universities under Grant 2021FZZX001-20.}
}

\author{\IEEEauthorblockN{Zhenguo Zhang, Qianqian Yang$^{\dag}$, Shibo He, Jiming Chen}
\IEEEauthorblockA{{
}
{College of Control Science and Engineering, Zhejiang University, Hangzhou 310007, China}\\
{College of Information Science and Electronic Engineering, Zhejiang University, Hangzhou 310007, China}\\
{The State Key Laboratory of Industrial Control Technology, Hangzhou 310007, China}\\
Key Laboratory of Collaborative Sensing and Autonomous Unmanned Systems of Zhejiang Province,\\
Zhejiang University, Hangzhou 310007, China\\
\{zhangzhenguo, qianqianyang20$^{\dag}$, s18he, 
  cjm\}@zju.edu.cn
  }

}
\maketitle

\begin{abstract}

Semantic communication has emerged as a promising approach for improving efficient transmission in the next generation of wireless networks. Inspired by the success of semantic communication in different areas, we aim to provide a new semantic communication scheme from the semantic level. In this paper, we propose a novel DL-based semantic communication system for video transmission, which compacts semantic-related information to improve transmission efficiency. In particular, we utilize the Bi-optical flow to estimate residual information of inter-frame details. We also propose a feature choice module and a feature fusion module to drop semantically redundant features while paying more attention to the important semantic-related content. We employ a frame prediction module to reconstruct semantic features of the prediction frame from the received signal at the receiver. To enhance the system's robustness, we propose a noise attention module that assigns different importance weights to the extracted features. Simulation results indicate that our proposed method outperforms existing approaches in terms of transmission efficiency, achieving about 33.3\% reduction in the number of transmitted symbols while improving the peak signal-to-noise ratio (PSNR) performance by an average of 0.56dB.



\end{abstract}

\section{Introduction}

The rapid development of internet applications, such as Internet-of-Things, has led to a significant increase in the scale of wireless video data, resulting in higher wireless communication demands{\cite{ng2022context}}. The conventional approach to wireless video transmission involves separation-based digital communication. It consists of source codecs (i.e., AVC/H.264{\cite{wiegand2003overview}}, HEVC/H.265{\cite{sullivan2012overview}}) and channel codecs (i.e., LDPC, Turbo), which have achieved impressive performance on wireless video transmission. However, the ``cliff-effect'' is the inevitable shade of the traditional separation-based wireless communication schemes over the physical channels{\cite{bourtsoulatze2019deep}}. That is, the decoder is unable to reconstruct the original information and instead produces random data when the channel condition falls below a certain threshold. 
Additionally, conventional or semi-DL-based video transmission systems fail to jointly design the transmitter and receiver, which hinders transmission efficiency.


Different from separation-based digital communication systems, semantic communication methods prioritize the meaning behind bits rather than solely focusing on optimizing bit-error rate (BER). These systems aim to reconstruct the original information or perform goal-oriented intelligent tasks at the receiver. DL plays a critical role in extracting semantic-related information and resisting channel noise{\cite{xie2021deep}}, and these systems
have shown exceptional performance in the transmission of text{\cite{xie2020lite}}, speech{\cite{han2022semantic}}, image{\cite{zhang2022semantic}}, and video{\cite{tung2022deepwive}}. In{\cite{xie2020lite}}, the authors proposed a text semantic communication system for Internet-of-Things devices, utilizing channel state information (CSI) aided scheme to achieve remarkable performance on different physical channels. {\cite{han2022semantic}} proposed a semantic communication method for speech transmission, where soft alignment and redundancy removal modules are used to extract semantic-related features and throw away semantic irrelevant ones. The author in{\cite{zhang2022semantic}} introduced an end-to-end multi-task image-related semantic communication system, presenting a novel approach for single-model data to achieve multiple tasks with different semantic information.

\begin{figure*}[htbp]
\centering
\subfloat[the proposed video semantic communication system]{\includegraphics[width=4.8in]{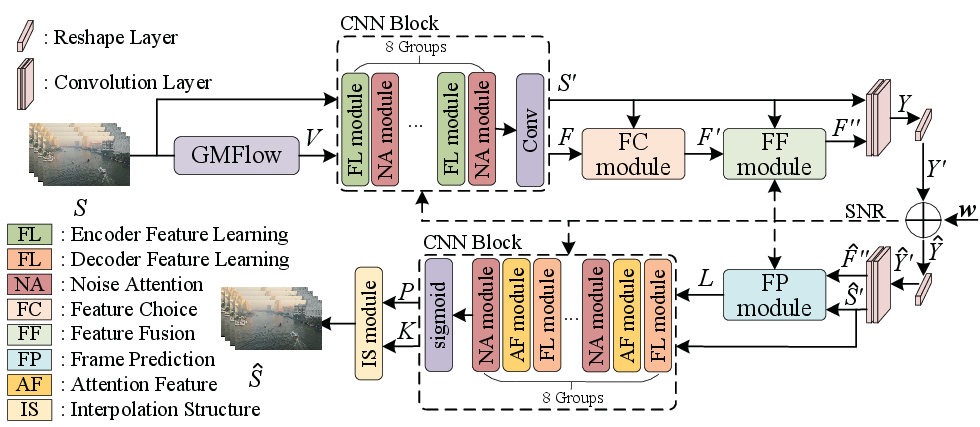}%
\label{fig2a}}
\hfil
\subfloat[encoder/decoder CNN Block]{\includegraphics[width=2.1in]{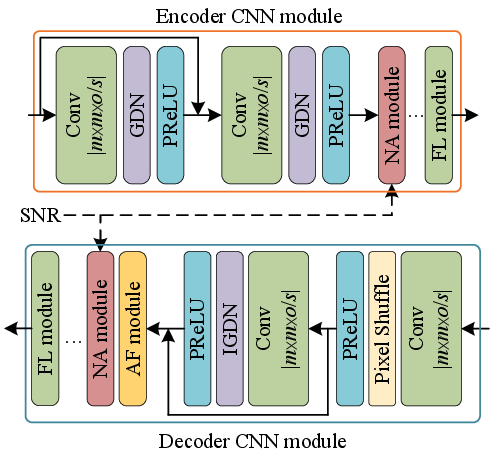}%
\label{fig2b}}
\caption{The proposed video semantic communication system and the CNN block.}
\label{fig:2}
\end{figure*}



For video transmission, both traditional and semantic communication systems rely on residual information from inter-frame details, which can be estimated using motion information{\cite{liu2021deep}} or optical flow{\cite{wu2020foreground}}. 
A joint source and channel coding (JSCC) based video transmission scheme has been proposed in{\cite{tung2022deepwive}}, which achieves better performance then traditional method by utilizing reinforcement learning (RL) to allocate the bandwidth for different video frames dynamically. The authors in{\cite{wang2022wireless}} have also proposed an end-to-end wireless video transmission scheme, named DVST. The system employs nonlinear transform and conditional coding architecture to extract semantic information from the inter-frame details. 
In{\cite{jiang2022wireless}}, a wireless semantic communication system has been established for video conferencing, which exploits an incremental redundancy hybrid automatic repeat-request architecture under different channel conditions (SVC-HARQ), dramatically reducing transmission resources.
However, these works do not explore the influence of the residual information estimation accuracy for video reconstruction at the receiver.

Inspired by the successful application of bidirectional flow to occlusion detection {\cite{xu2022gmflow}}, we propose an innovative bidirectional flow based semantic communication system for wireless video transmission. The contributions of this paper can be summarized as follows:

\begin{itemize}
\item{ A DL-based video semantic communication system with the Bi-optical flow extractor is proposed to extract more fine-grained semantic inter-frame information. To the best of our knowledge, we are the first to exploit Bi-optical flow to enhance video transmission efficiency in the semantic communications literature.}

\item{We exploit the noise attention module to mitigate the influence of the imperfect channel conditions, which achieves a graceful performance degradation when the channel quality drops below the desired signal-to-noise (SNR) value.}

\item{
We propose a feature choice module and a feature fusion module to drop semantically redundant features and pay more attention to the crucial semantic latent features. We also employ a frame prediction module to reconstruct the video frame at the receiver.

}

\item{The numerical results validate the effectiveness of our approach, which saves about 33.3\% of the transmission length for achieving the equivalent or slightly better performance.}

\end{itemize}

The rest of the paper is organized as follows. In Section \textrm{II}, we introduce the system model. The details of the proposed architecture are shown in \textrm{III}. Section \textrm{IV} presents the numerical results, and Section \textrm{V} concludes the paper.

\section{Proposed Method}

\begin{figure*}[htbp]
\centering
\subfloat[feature fusion module]{\includegraphics[width=4in]{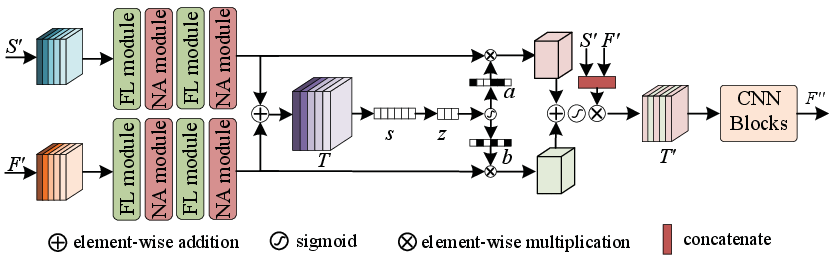}%
\label{fig6a}}
\hfil
\subfloat[noise attention module]{\includegraphics[width=2.3in]{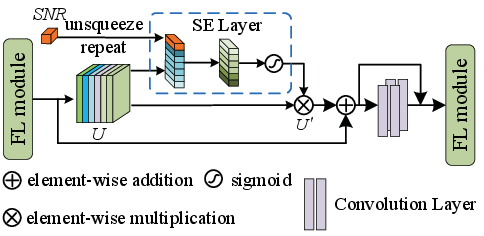}%
\label{fig6b}}
\caption{The proposed feature fusion module and the noise attention module.}
\label{fig:4}
\end{figure*}

In this section, we introduce our proposed DL-based semantic communication system for wireless video transmission, which is depicted in Fig. 1(a). The considered semantic communication system comprises several key components that work together to improve semantic accuracy and enhance robustness. Specifically, the GMFlow model{\cite{xu2022gmflow}} extracts Bi-optical flow information, while the convolutional neural network (CNN) block progressively downsamples input features. The feature choice and fusion modules select and learn essential features to enhance semantic accuracy. The channel encoder and decoder pay more attention to semantic-related features to improve the system's robustness. The frame prediction module reconstructs compressed critical features, which are then upsampled to their original dimensions using the Pixel Shuffle module in the CNN block. Finally, the interpolation structure recovers the original video frame sequence. In the following subsections, we will provide a detailed description of each component of our proposed system.

\subsection{Semantic Encoder}

We propose a semantic communication system for video transmission, depicted in Fig. 1(a), which consists of a transmitter and receiver. The video sequence is denoted as $\{\bm{s}_i^n\}$, where $s_i^n\in\mathfrak{R}^{H\times{W}\times{C}}$, $\forall{i}\in[1,N]$ represents the $i$-th video frame in the $n$-th group of $N$ sequential frames. $H$ and $W$ are the height and width of the video frame, respectively. $C$ is the number of channels. The goal of video transmission is to reconstruct subsequent video clip $\hat{s}_i^n$ at the receiver based on the received symbols. In accordance with the literature on JSCC literature{\cite{bourtsoulatze2019deep}}, we denote the channel output size of the encoder by $k$ and the size of each video frame by $m$. The compression ratio of the bandwidth is $\rho=\frac{k}{m}=\frac{k}{3HWN}$.

The semantic encoder is illustrated in Fig. 1(a), which extracts latent representation from the source information. It consists of four components, namely GMFlow, a CNN block, a feature choice module, and a feature fusion module. The GMFlow module{\cite{xu2022gmflow}} first captures the dense features $\bm{F_1, F_2}\in\mathfrak{R}^{H\times{W}\times{D}}$ from sequential video frames using weight-sharing convolutional networks, where $D$ is the feature dimension. Then, it investigates the correspondences of the sequential video frames with pixel features in $\bm{F_1}$ and $\bm{F_2}$ by computing their correlations
{\cite{wang2020learning}}
$\bm{C}=\frac{\bm{F_1F_2^T}}{\sqrt{D}}$, where $\bm{C}$ is the correlation value, and $\frac{1}{\sqrt{D}}$ is a normalization factor to limit the matrix dot-product result. The correspondence $\bm{\hat{G}}$ is obtained by $\hat{\bm{G}}={\rm{softmax}}(\bm{C})\bm{G}$, where $\bm{G}\in\mathfrak{R}^{H\times{W}\times{2}}$ is the weighted average of the pixel grid. Finally, The difference of the correspondences pixel gives the optical flow with bidirectional $\bm{V}=\hat{\bm{G}}-\bm{G}\in\mathfrak{R}^{H\times{W}\times{4}}$. We use the bidirectional optical flow to estimate the residual information of consecutive frames, which improves the accuracy of residual estimation.

Fig. 1(b) illustrates the architectures of the feature extraction module used for learning information. The encoder CNN module extracts the semantic information by convolutional layers with \textit{m × m × o/s}, where \textit{m} is the kernel size, \textit{o} is the number of kernels, and \textit{s} is the stride. Additionally, a generalized normalization transformation layer (GDN/IGDN) is applied to each convolutional layer to generalized divisive normalization {\cite{balle2015density}}. The output of all convolutional layers is nonlinear mapped using the PReLU activation function (sigmoid nonlinearity in the receiver CNN block). The pixelshuffle is used as the upscale module.

The noise attention module following each feature learning module redistributes the weight of the learned latent representation to mitigate the effect of channel noises, as shown in Fig. 2(b). The SNR value first extends to the same dimension as the latent representation processed by the average pooling function. Then the concatenated result is fed into the squeeze and excitation (SE) module to obtain the output weight. The scaled features $\bm{U^{'}}$ are obtained by multiplying the input features with the scaling factor. Finally, we use the residual network to learn the noise effect on the system. The noise attention module is embedded into different modules to improve the system robustness for the channel noise.

\begin{figure}[htbp]
\centering
\includegraphics[scale=0.63]{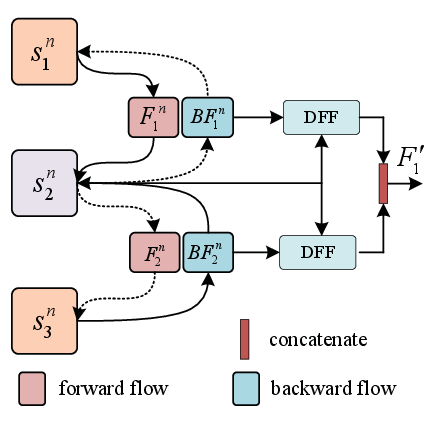}
\caption{The proposed feature choice module.}
\label{fig:7}
\end{figure}

We obtain the optical flow $\bm{F}\in\mathfrak{R}^{\frac{H}{t}\times{\frac{W}{t}}\times{512}}$ and video frames $\bm{S^{'}}\in\mathfrak{R}^{\frac{H}{t}\times{\frac{W}{t}}\times{128}}$ latent representation from the encoder feature learning module, where $t$ represents the downsampling factor. The substantial amount of input semantic information brings enormous challenges to the transmitter for paying different attention to abundant features. To address this issue, we propose a feature choice module to refine the semantic information while reducing complexity. In this module, we keep the optical flow and key frame information denoted as the solid line and discard the semantic information indicated by the dashed line in Fig. 3. The dual features fusion (DFF){\cite{zhang2022semantic}} method is employed to fuse frame and optical flow information. 

For the complex inter-frame information in the spatial and temporal domain of the video transmission over the physical channel, we propose a feature fusion module based on{\cite{li2019selective}} to concentrate on the critical semantic features for better reconstruction performance. The generated feature $\bm{F^{'}}\in\mathfrak{R}^{\frac{H}{t}\times{\frac{W}{t}}\times{256}}$ from the feature choice module and $\bm{{S^{'}}}$ are fed into the feature fusion module, which outputs $\bm{F^{''}}\in\mathfrak{R}^{\frac{H}{t}\times{\frac{W}{t}}\times{256}}$. As depicted in Fig. 2(a), $\bm{{F^{'}}}$ and $\bm{{S^{'}}}$ are processed by the feature learning and noise attention module. This module uses different kernel sizes to achieve various receptive fields to improve the transmission efficiency of the latent semantic representation in a soft-attention manner. The module utilizes a more suitable method to update the weight of semantic information.

\subsection{Channel Encoder and Decoder}

The channel encoder and decoder consist of convolution and reshape layers. The cascaded convolution layers of the channel encoder map the features $\bm{S^{'}}$ and $\bm{F^{''}}$ to $\bm{Y}={\{\bm{Y_1, Y_2}\}}\in\mathfrak{R}^{\frac{H}{t}\times\frac{W}{t}\times({y_1,y_2})}$, which reshapes into $\bm{Y^{'}}={\{\bm{Y}_{1}^{'},\bm{Y}_{2}^{'}\}}\in\mathfrak{R}^{c{'}\times{2}}$, where $c{'}$ is $\frac{1}{2}\times{\frac{H}{t}\times{\frac{W}{t}}\times{(y{_1},y{_2}})}$, $y{_1}$ and $y{_2}$ are the channel output size of the key and prediction frame features. The front and back channels correspond to the real and imaginary parts of the wireless symbols to be transmitted, respectively. At the receiver, the symbol sequences $\bm{\hat{Y}}$ are reshaped into $\bm{\hat{Y}^{'}}={\{\bm{\hat{Y}_{1}^{'}, \hat{Y}_{2}^{'}}\}}$, which has the same dimension as $\bm{Y}$. The semantic features $\bm{\hat{Y}^{'}}$ are then fed into cascaded convolution layers to recover the video-related latent representation $\bm{\hat{S}^{'}}$ and $\bm{\hat{F}^{''}}$. 


\subsection{Semantic Decoder}

The semantic decoder is responsible for mapping the latent representation to recover the original video signal. In this work, we propose a frame prediction module inspired by {\cite{xiao2018weighted}}, which leverages the compressed semantic features to reconstruct the future frame. The architecture first concatenates the received symbols $\bm{\hat{F}^{''}}\in\mathfrak{R}^{\frac{H}{t}\times{\frac{W}{t}}\times{256}}$ and $\bm{\hat{S}^{'}}\in\mathfrak{R}^{\frac{H}{t}\times{\frac{W}{t}}\times{128}}$. The resulting features are then processed by a simplified Res-UNet module to obtain the semantic features $\bm{L}\in\mathfrak{R}^{\frac{H}{t}\times{\frac{W}{t}}\times{128}}$. 
The simplified Res-UNet progressively downsamples the concatenated features by CNN layers, then upsamples it back to the original dimensions, as shown in Fig. 4.
Meanwhile, we embed the noise attention module in the upsampling section to mitigate the effects of the channel noise. The CNN block maps the learned features $\bm{L}$ and $\bm{\hat{S}^{'}}$ to $\bm{P}$ and $\bm{K}$ with the same dimension of $\bm{S}$. We inset the Pixel Shuffle module in the decoder feature learning module to increase the feature dimensions by exchanging channel dimensions for height and width dimensions. To enhance the expressive capacity of the system, we use the attention feature (AF) module proposed in{\cite{fu2021learned}}, which gives a more exquisite description of the important areas. Finally, an interpolation structure is employed to recover the original sequence of video frames.

\begin{figure}[htbp]
\centering
\includegraphics[scale=0.76]{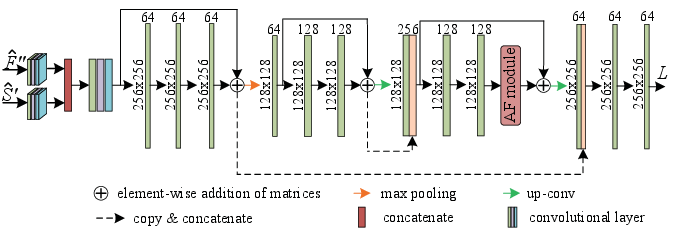}
\caption{Architecture of the frame prediction module.}
\label{fig7}
\end{figure}

\section{Numerical results}

In this section, we evaluate the performance of the proposed semantic communication system with the existing video transmission schemes. We train the proposed method with the UCF101 dataset{\cite{soomro2012ucf101}}, which contains 13320 videos of 101 semantic labels with the resolution of 320 × 240. We evaluate different schemes on the BVI-DVC dataset{\cite{ma2021bvi}}. The dataset consists of 800 video sequences ranging from 270p to 2160p.
We set the learning rate to 0.0001, the batch size to 8, and the downsampling factor \textit{t} = 8, respectively. The estimated $\rm{SNR_{est}}$ at the transmitter and receiver is $10\log_{10}{\frac{T}{\sigma^2}}$, where ${T}$ is the transmitter power constraint. To train our model, we use dynamically varied SNR values, randomly sampling SNR from -5dB to 15dB in each batch. We compare the performance of our proposed method with the existing semantic communication approach (\textit{DeepWiVe}){\cite{tung2022deepwive}} and separation-based traditional schemes, using H.264 for source coding, LDPC codes for channel coding, and QAM for modulation. We employ the PSNR and multi-scale structural similarity index measure (MS-SSIM){\cite{wang2004image}} to evaluate the quality of the reconstructed video frames at the receiver.


\subsection{Simulation Results}

Fig. 5 presents the performance comparison of the proposed semantic communication system with traditional separation-based digital schemes and the DL-based approach, \textit{DeepWiVe}. The results are shown in terms of PSNR and MS-SSIM metrics in Fig. 5(a) and 5(b), respectively. We can observe that our proposed method achieves better performance than the benchmark on different metrics with compression ratio $\rho=0.031$. Specifically, our approach achieves an average improvement of 6.76dB in PSNR and 0.084 in MS-SSIM over the traditional methods under the AWGN channel. When compared to \textit{DeepWiVe}, our method achieves a gain of 2.09dB in PSNR and 0.0079 in MS-SSIM. 
We further reduce the number of the transmitted symbols to evaluate the system performance with $\rho=0.021$, which saves 33.3\% of wireless data traffic. Our proposed scheme still outperforms \textit{DeepWiVe} by 0.56dB in PSNR. However, \textit{DeepWiVe} has better structural metric performance.  From the figure, we observe that our method has a slight advantage over \textit{DeepWiVe} at moderate channel conditions. This is due to that the \textit{DeepWiVe} could reasonably distribute the bandwidth to the key and remaining frames using reinforcement learning. 
Moreover, the system performance is mainly determined by the bandwidth compression ratio when the channel conditions very better. Compared with benchmark schemes, our proposed method presents a better capacity for video frame reconstruction.

\begin{figure}[htbp]
\centering
\subfloat[PSNR]{\includegraphics[width=3.1in]{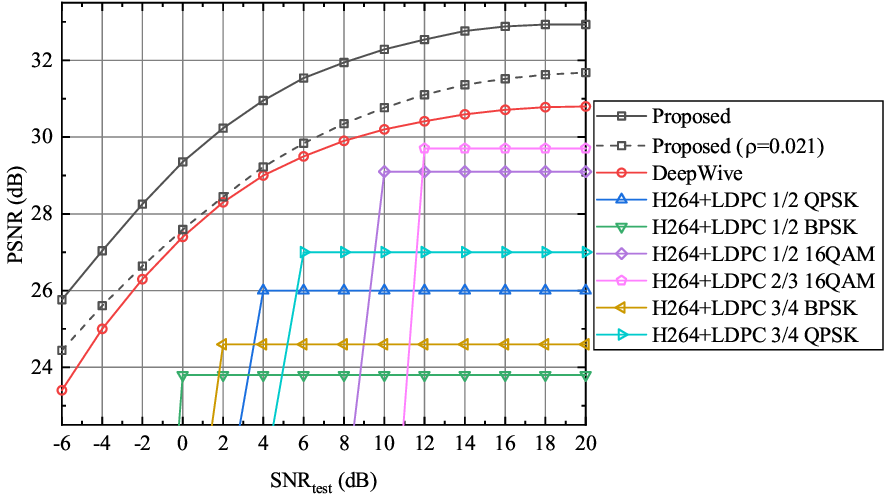}}
\label{fig8a}

\subfloat[MS-SSIM]{\includegraphics[width=3.1in]{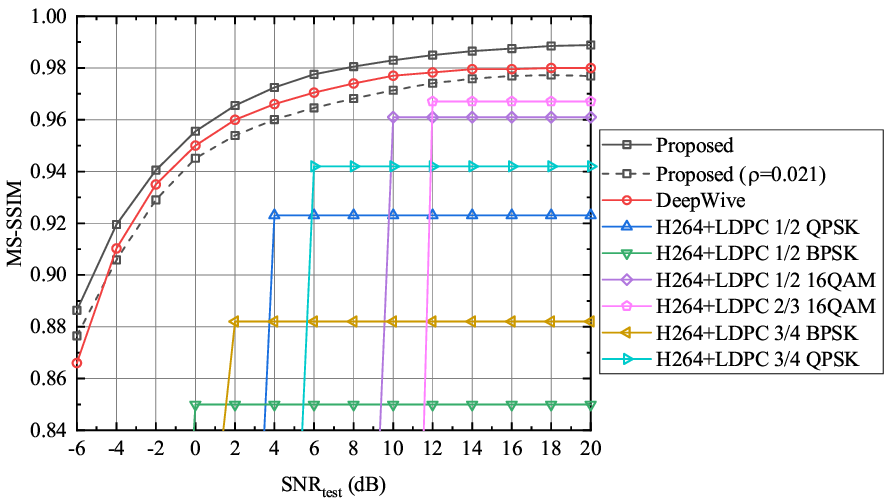}%
\quad
\label{fig8d}}

\caption{Performance comparison of different approaches over AWGN channel for bandwidth compression ratio $\rho=0.031$.}
\label{fig8}
\end{figure}

\subsection{Ablation Study}

The communication system with the noise attention module relies on accurate estimation of channel conditions at both the transmitter and receiver. We evaluate the performance of different schemes under various $\rm{SNR_{est}}$ values, as depicted in Fig. 6. Our DL-based communication system outperforms traditional methods for all $\rm{SRN_{test}}$ values, regardless of the $\rm{SNR_{est}}$. Compared to \textit{DeepWiVe}, our proposed method shows a more graceful curve variation for different channel estimations, demonstrating the robustness of our end-to-end system. 
The noise attention module not only mitigates the effect of channel noise, but also improves the resilience of the transmitter and receiver to variations in the estimated channel conditions. Moreover, the DL-based semantic communication system does not suffer from the ``cliff effect'' associated with conventional digital schemes, which randomly reconstruct the source information in terrible channel conditions. This behavior illustrates that the DL-based semantic communication system bridges an implicit trade-off between communication reliability and compression ratio.

\begin{figure}[htbp]
\centering
\includegraphics[width=3.3in]{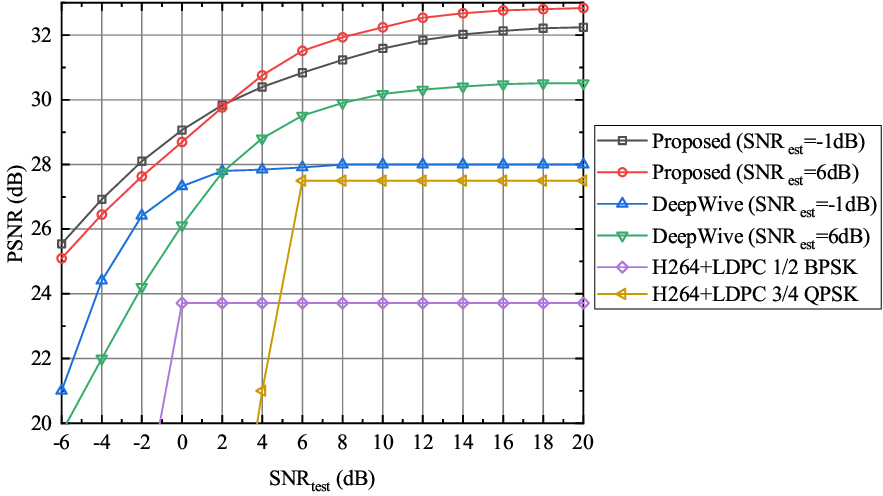}
\caption{Performance of different approaches with respect to the $\rm{SNR_{est}}$ over an AWGN channel for bandwidth compression ratios $\rho=0.031$.}
\label{fig:10}
\end{figure}



\section{Conclusion}

In this paper, we proposed a novel DL-based semantic communication system for video transmission. The Bi-optical flow is utilized to estimate the residual information of the consecutive frames. We design a feature choice module and a feature fusion module to drop semantically redundant features and pay more attention to the crucial semantic latent features. Additionally, the frame prediction module excavates the compressed latent semantic representation received from the transmitter to improve system performance. To mitigate the effect of the physical channel, we propose a novel noise attention module that redistributes learned feature weights. In particular, the system overcomes the ``cliff effect'' and exhibits graceful performance degradation for different channel conditions. Experiment results show that our approach outperforms benchmark methods with various metrics by dropping semantically irrelevant features and improving video reconstruction quality.

\bibliographystyle{ieeetr}
\bibliography{conference}

\end{document}